\documentclass[a4paper,10pt]{article}
\usepackage[super,sort&compress]{natbib}
\usepackage[noblocks]{authblk}
\usepackage{amsmath}
\usepackage{pifont}
\usepackage{subfigure}
\usepackage{color}
\usepackage{ulem}
\usepackage{graphicx}
\usepackage{multicol}
\usepackage{bbding}
\usepackage{url}
\newcommand{\cmark}{\scriptsize\Checkmark \normalsize}
\newcommand{\xmark}{\scriptsize \XSolidBrush \normalsize}

\begin{document}

\title{Activity clocks: spreading dynamics on temporal networks of human contact}
\author{Laetitia Gauvin}
\author{Andr\'e Panisson}
\author{Ciro Cattuto \footnote{ciro.cattuto@isi.it}}
\affil{{\small Data Science Lab, ISI Foundation,
Torino, Italy}}
\author{Alain Barrat}
\affil{{\small Aix Marseille Universit\'e, CNRS, CPT, UMR 7332, 13288 Marseille, France\\ 
Universit\'e de Toulon, CNRS, CPT, UMR 7332, 83957 La Garde, France\\
Data Science Lab, ISI Foundation,
Torino, Italy}}

\maketitle

\begin{abstract}
Dynamical processes on time-varying complex networks are key to understanding and modeling a broad variety of processes in socio-technical systems. 
Here we focus on empirical temporal networks of human proximity and we aim at understanding the factors that, in simulation, shape the arrival time distribution of simple spreading processes.
Abandoning the notion of wall-clock time in favour of node-specific clocks based on activity exposes robust statistical patterns in the arrival times across different social contexts. Using randomization 
strategies and generative models constrained by data, we show that these patterns can be understood in terms of heterogeneous inter-event time distributions coupled with heterogeneous numbers of events per edge.
We also show, both empirically and by using a synthetic dataset, 
that significant deviations from the above behavior can be caused by the presence of edge classes with strong activity correlations.
\end{abstract}

The field of complex networks has recently undergone an important evolution.
Thanks to the recent availability of time-resolved data sources,
many studies performed under the assumption of static network structures
can now be extended to take into account the network's dynamics.
Data on time-varying networks are becoming accessible
across a variety of contexts,
ranging from communication networks~\cite{Eckmann:2004,Holme:2005,Onnela:2007,Makse:2009,Amaral:2009,Karsai:2012}
to proximity networks~\cite{Cattuto2010,Salathe:2010}
and infrastructure networks~\cite{Gautreau:2009,Bajardi:2011}.
This avalanche of data is prompting a surge of activity
in the field of ``temporal networks''~\cite{Review-TNW-Holme}.
Data analysis has shown the coexistence of statistically stationary properties
and topological changes, as well as the burstiness of interactions
characterized by highly skewed distributions of inter-event times~\cite{Eckmann:2004,Holme:2005,Onnela:2007,Makse:2009,Amaral:2009,Gautreau:2009,Cattuto2010,Salathe:2010,Bajardi:2011,Karsai:2012,humdyn-barabasi,humdyn-vazquez,Barabasi:2010,Review-TNW-Holme}.
These temporal features of networks influence the dynamics of network processes,
just like the topological structure of static networks does~\cite{Barrat:2008}.
As a consequence, and  similarly to the case of static networks,
simple dynamical processes such as random walks~\cite{Starnini}, synchronization phenomena \cite{Prignano:2012}, 
consensus formation~\cite{Baronchelli:2012}
or spreading processes~\cite{Vazquez:2007,Iribarren:2009,Miritello:2011,Isella:2011,karsai-2011-83,Kivela2012,Rumor-CNW-Moreno04}
can be used as probes to investigate the temporal and structural properties of time-varying networks.

Previous works on the dynamics of spreading processes over complex networks
have considered both the topological and the temporal structure of networks~\cite{karsai-2011-83,Vazquez:2007,Review-TNW-Holme},
as well as the specific impact that the temporal structure bears on the spreading process.
The quantities used to quantify the measured effects are typically network averages,
such as the outbreak sizes of an epidemic or its prevalence.
These average quantities, however, fail to account for important heterogeneities
in the arrival times of the spreading process.
Recent work~\cite{Panisson2011} showed that the non-stationarity and burstiness
of empirical temporal networks lead to noisy distributions of arrival times,
and that shifting the perspective from a global notion of wall-clock time
to a node-specific ``time'' based on node activity allows to expose
a clear and robust pattern in arrival ``times''.

Here we focus on the distribution of arrival times for spreading processes,
based on a wide range of empirical data on time-resolved human proximity.
In particular, we seek to identify the dynamic features of the temporal network
that are responsible for the observed arrival time distributions.
To this aim, we consider temporal networks of human contacts
and we define hierarchies of null models and generative models
that selectively retain or discard specific properties of the empirical data.
We simulate simple spreading processes over these models
and perform a comparative analysis of the arrival time distributions.
Our results identify the most salient properties that characterize realistic models of human interaction networks,
and highlight the properties that control the arrival time distributions,
with applications to several domains such as opportunistic information transmission
and epidemic spread and containment.

%
%
%
We consider time-varying networks of human proximity measured using wearable sensors.
The data were collected by the SocioPatterns collaboration (\url{http://www.sociopatterns.org})
in different social contexts:
two conferences in Italy (HT09) and France (SFHH) \cite{Cattuto2010,Isella:2011},
a primary school in France (PS) \cite{Stehle:2011},
and a paediatric hospital ward in Italy (HOSP) \cite{Isella:2011b}.
Details on the data collection methods are reported in the Supplementary Information description and Table $S1$.
All of the datasets we consider describe the face-to-face proximity relations
of the monitored subjects, with a temporal resolution of approximately $20$ seconds~\cite{Cattuto2010,Isella:2011}.
For every pair of individuals, the full sequence of individual interactions is resolved,
with starting and final timestamps for every close-range proximity relation.
These data can be represented as time-varying networks of proximity:
nodes represent individuals and a link connecting two nodes
indicates that the corresponding individuals are in contact, i.e.,
in face-to-face proximity of one another.

\section*{Results}
\subsection*{\small Epidemic Processes and Activity Clocks}
We probe the temporal structure of the empirical networks
with a simple Susceptible-Infected (SI) process.
The population of nodes (individuals) is split into two compartments:
susceptible nodes (S), who have not caught the ``infection'',
and infected nodes (I), who carry the ``infection'' and may propagate it to others.
In this simple epidemic model, infected nodes never recover.
A node is randomly selected as the seed from which the infection starts spreading deterministically,
through contacts between a susceptible node and an infected one ($S+I \rightarrow 2I$).
Transmission events are assumed to occur instantaneously on contact.

We fingerprint the temporal network structure of the data by computing the times at which
the epidemic process reaches the different nodes.
Specifically, we focus on the probability distribution of arrival times for the SI process unfolding over the temporal network.
In terms of wall-clock time, the arrival time at a given node is defined as the time elapsed
between the start (seeding) of the SI process and the time at which the process reaches the chosen node.
It has been shown\cite{Panisson2011} that the distribution of these arrival times
is extremely sensitive to several heterogeneities of the empirical data,
to the seeding time. In general, it displays strong heterogeneities
due to the non-stationary and bursty behavior of empirical temporal networks
that cannot be captured by simple statistical models.
Thus, we shift to a node-specific definition of ``time'':
each node is assigned its own ``activity clock'' that measures the time that node has spent in interaction
or, similarly, the number of contact interactions that node has been involved in.
The ``time'' measured by this clock does not increase when the node is isolated from the rest of the network.
In the following, for clarity, we will indicate with ``time$^*$'' the activity-clock readings.
The ``arrival time$^*$'' of the epidemic process at a given node is defined as the increase
of its activity clock reading from the moment the SI process is seeded to when it reaches the node.
Arrival times$^*$ discard by definition many temporal heterogeneities
of the empirical data and usually exhibit a well-defined distribution~\cite{Panisson2011}
that is robust with respect to changes in the starting time of the process
and across temporal networks of human contact measured in different contexts.
In the following we use activity clocks based on the number of contact events
a node has been involved in. The arrival time$^*$ at a node, consequently,
will be integer-valued and will measure the number of interactions each node was part of
from the seeding of the epidemic until the node was infected.

For each empirical time-varying network,
we generate a hierarchy of synthetic temporal networks using both a top-down and a bottom-up approach.
The synthetic networks are designed to support our analysis by selectively retaining or discarding
specific properties of the empirical data.

\subsection*{\small Top-down approach: Null models}
We generate null models by applying to the empirical data randomization procedures that erase specific correlations~\cite{Kivela2012}.
We keep the topology of the contact network unchanged.
In the ``interval shuffling'' (IS) procedure, the sequences of contact and inter-contact durations
are reshuffled for each link separately,
while in the ``link shuffling'' (LS) procedure~\cite{Kivela2012} the unaltered sequences
of events are swapped between link pairs.
Both procedures destroy the causal structure of the temporal network,
but they both preserve the global distributions of contact durations,
inter-contact durations, and number of contacts.
The IS procedure also preserves, for every link,
the total number of contact events and the cumulated interaction time,
while the LS procedure does not conserve these quantities at the link level.

We also consider a global time shuffling procedure (TS):
we build a global list of the empirical contact durations
and, for each link, we generate a synthetic activity timeline
by sampling with replacement the global list of contact durations
according to the original number of contacts for that link.
While the global distribution of contact durations and of the number of contacts per link
are conserved by construction, all temporal correlations are destroyed
and the distributions of inter-contact times differs from the empirical one.

\begin{figure}[!htbp]
\begin{center}
\includegraphics[width=\columnwidth] {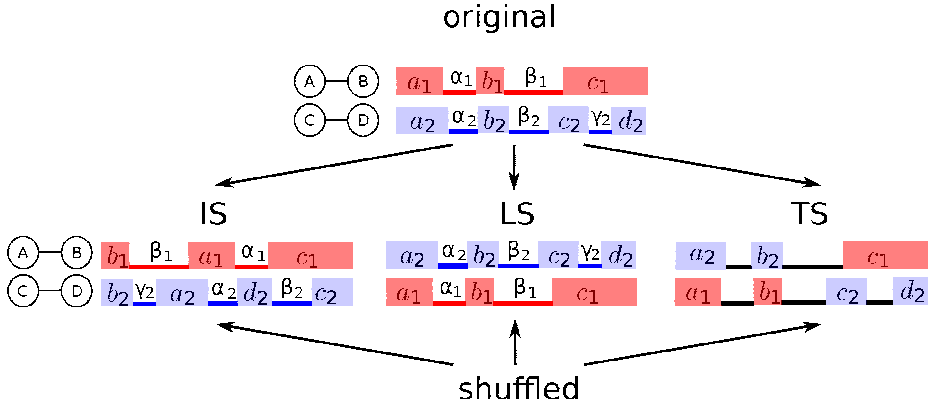}
\caption{Example of the shuffling procedures for the simple case of a network with four nodes (A, B, C, D) and  two links (A--B and C--D) with their respective contact sequences. Red (light) segments indicate A--B contacts, while blue (dark) segments indicate C--D contacts.
For each link, individual contact intervals are marked with latin letters and inter-contact intervals with greek letters.
IS, LS, and TS stand, respectively, for Interval Shuffling, Link Shuffling and Time Shuffling.
In the TS case the inter-contact intervals are determined by the sampled contact intervals and do not correspond to inter-contact intervals of the original data.
\label{fig:shuffling}
}
\end{center}
\end{figure}

Figure~\ref{fig:shuffling} illustrates the three randomization procedures defined above.
All the procedures conserve the topology, 
the distribution of contact durations and the distribution of the number of contacts per link 
of the empirical networks. Table~\ref{Table_Mod1} summarizes 
the impact of the randomization procedures on different properties of the temporal networks.

\begin{table}[!htbp]
\centering
\begin{tabular}{|l|c|c|c|c|c|c|c|c|}
\hline
Models & topology & causality & $P(\tau)$ & $\omega_{AB}$ & $P(\omega)$ & $n_{AB}$& $P(n)$  \\
\hline
IS   &\cmark  &\xmark&\cmark&\cmark&\cmark&\cmark&\cmark\\
LS  &\cmark  &  \xmark&\cmark&\xmark&\cmark&\xmark&\cmark\\
TS  &\cmark   & \xmark&\xmark&\xmark&\xmark&\cmark&\cmark\\
\hline
\end{tabular}
\caption{Properties of the empirical temporal networks that are retained (\cmark) or discarded (\xmark) 
by the various null models.
$P(\tau)$ is the distribution of inter-contact interval durations.
$\omega_{AB}$ indicates the cumulated contact durations of an arbitrary link $AB$,
and $P(\omega)$ is the distribution of cumulated contact durations.
$n_{AB}$ indicates the number of contacts per link of an arbitrary link $AB$,
and $P(n)$ is the distribution of the number of contacts per link.
IS, LS, and TS stand, respectively, for Interval Shuffling, Link Shuffling and Time Shuffling.
\label{Table_Mod1}
}
\end{table}

\subsection*{\small Bottom-up approach: Generative models}
 We also define generative models for random temporal networks
designed so that the resulting time-varying networks exhibit specific properties of the empirical data,
in the spirit of the configuration model for static networks~\cite{Catanzaro:2005}.
We start by creating a static random Erd\"{o}s-R\'{e}nyi network with the same number of nodes
and the same average degree of the temporally-aggregated empirical contact network. 
Then we assign to each link a sequence of synthetic contact events, according to different strategies.
In the Inter-Contact Time model (ICT) we impose
that the global distribution of inter-contact durations is the same as in the empirical data
(see the Methods section for details). This is an important case to test against,
as it is often considered in the literature that the distribution of inter-contact times
plays an important role in determining and constraining spreading processes over temporal networks~\cite{Panisson2011}. 
Contact durations are fixed and equal to the average contact duration measured in the empirical data.
In the Inter-Contact Time plus Contact-Per-Link model (ICT+CPL)
we proceed as in the ICT case, but also impose that the distribution of the number of contact events per link
must match the empirical one.
In summary, in both models the topology and the contact duration distribution differ from the empirical ones.
Table~\ref{Table_Mod} summarizes the properties of the generated temporal networks
that are constrained to match those of the empirical data.
\begin{table}[!htbp]
\centering
\begin{tabular}{|l|c|c|c|c|c|c|c|c|}
\hline
Models & topology & causality & $P(\tau)$ & $\omega_{AB}$ & $P(\omega)$ & $n_{AB}$ & $P(n)$ \\
\hline
ICT   & \xmark & \xmark &\cmark&\xmark&\xmark&\xmark&\xmark\\
ICT+CPL & \xmark & \xmark &\cmark&\xmark&\xmark&\xmark&\cmark\\
\hline
\end{tabular}
\caption{
Properties of the empirical temporal networks that are retained (\cmark) or discarded (\xmark)  by the generative models.
As in Table~\ref{Table_Mod1},
$P(\tau)$ is the distribution of inter-contact interval durations.
$\omega_{AB}$ indicates the cumulated contact durations of individual links,
and $P(\omega)$ is the distribution of cumulated contact durations.
$n_{AB}$ indicates the number of contacts per link,
and $P(n)$ is the distribution of the number of contacts per link.
ICT and ICT+CPL  stand, respectively, for the Inter-Contact Time model
and the Inter-Contact Time plus Contacts-Per-Link model.
}
\label{Table_Mod}
\end{table}

\subsection*{\small Arrival times measured with activity clocks}
\begin{figure}[!htbp]
\centering
\includegraphics[width=\textwidth,keepaspectratio]{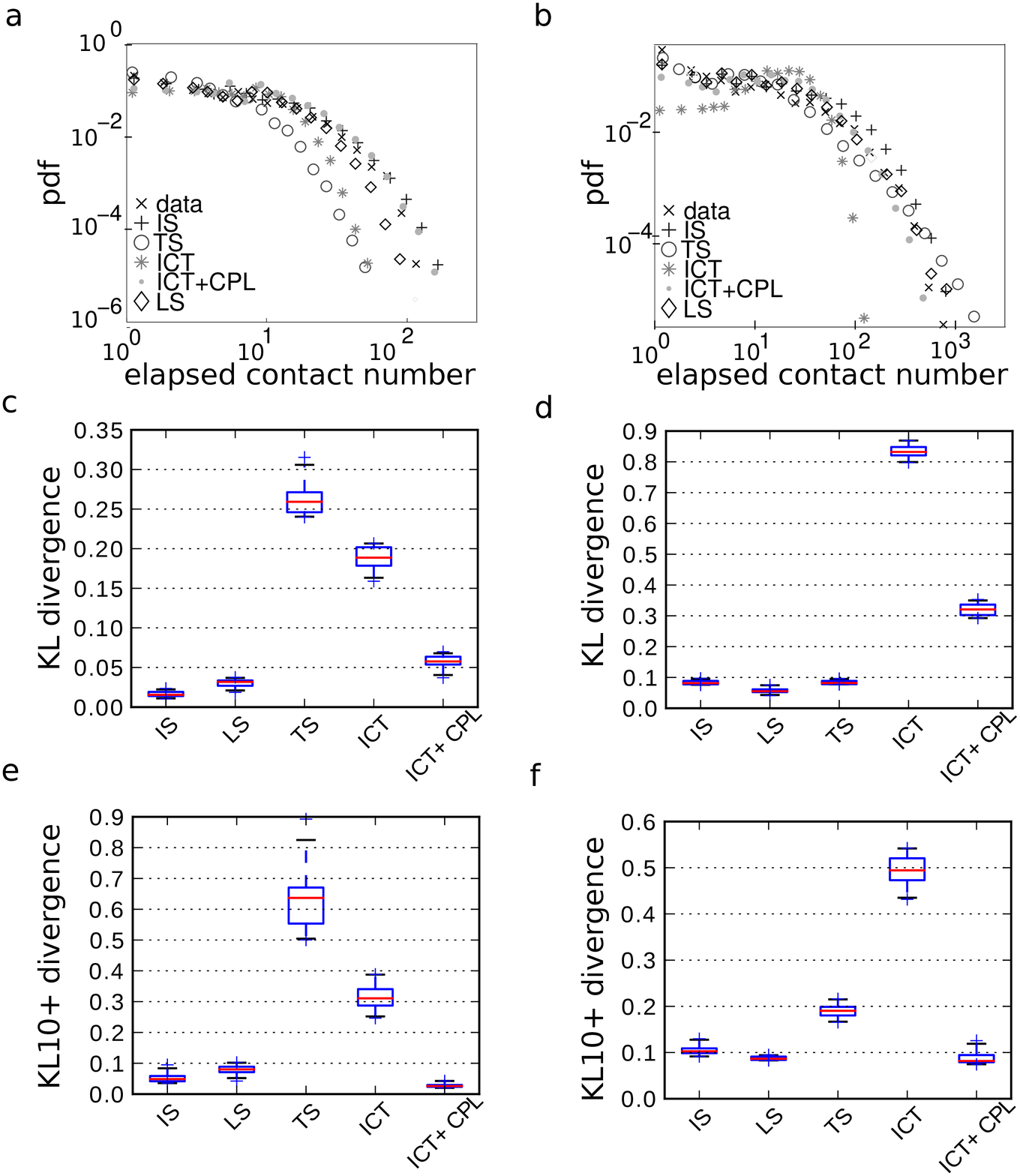}
\caption{Log-binned probability distributions (pdf) of arrival times$^*$  (top row) and
Kullback-Leibler divergences (middle row for KL and bottom row for KL10+)
for the conference dataset (HT09, left column) and the hospital data (HOSP, right column).
In each panel, IS, LS, TS, ICT, ICT+CPL stand respectively
for Interval Shuffling, Link Shuffling, Time Shuffling,
Inter-Contact time model, and Inter-Contact time plus Contacts-per-Link model.
In the top row, ``data'' indicates the distribution of arrival times$^*$
obtained by simulating an SI process over the empirical temporal network
($200$ realizations with random starting times for each node of the network taken as seed of the epidemics).
For each model, we consider $20$ different realizations of the temporal network.
For each of these realizations we run $20$ different SI epidemics, each with a different random starting time.
The arrival times$^*$ (top row) for all those runs are aggregated to yield the reported
distributions.
In the boxplots (middle and bottom row) the box extends from the lower to upper quartiles, and the line indicates the median value.
The whiskers of the box correspond to the $95 \%$ confidence interval.
\label{fig:composite}
}
\end{figure}
 From each empirical network, we build synthetic networks according to each null and generative model. 
We simulate SI processes on both empirical and synthetic networks for different starting times
and for different choices of the seed node. 
We then compute the distributions of arrival times$^*$ measured in terms of activity clocks.
Figure~\ref{fig:composite} (panels \textit{a} and \textit{b}) compares the arrival time$^*$ distributions
from the empirical data (HT09 conference and hospital datasets)
with those yielded by the null and generative models.
The results for the SFHH conference dataset are reported in the SI.

In order to provide a quantitative assessment of the distribution similarity
we compute the symmetrized Kullback-Leibler (KL) divergence\cite{Kullback1951} (see Methods)
between the distribution of arrival times$^*$ for the empirical data and for each model.
Given the relevance of large arrival time$^*$ values, which may be strongly influenced by causal constraints
and in general by the peculiarities of the temporal structure of the network,
we also compute the Kullback-Leibler divergence between the tails of the distributions.
To this end, we only take into account arrival times$^*$ longer than a fixed threshold
arbitrarily set to $10$. We refer to this restricted Kullback-Leibler divergence as ``KL10+'',
and we have checked that our results are robust with respect to changes in the threshold.
Table~\ref{tabKL} and Fig.~\ref{fig:composite} report the symmetrized KL and KL10+
divergences for the conference and hospital datasets we consider.
\begin{table}[!htbp]
\centering
\begin{tabular}{|l|l|l|l|l|l|l|l|l|}
\hline
& \multicolumn{2}{|c|}{HT09 conference} & \multicolumn{2}{|c|}{SFHH congress} & \multicolumn{2}{|c|}{hospital }\\
\hline
Models  & KL    & KL10+       & KL    & KL10+          & KL    & KL10+ \\
\hline
IS      & 0.012 & 0.032       & 0.011 & 0.031          & 0.067 & 0.079 \\
LS      & 0.022 & 0.052       & 0.023 & 0.085          & 0.053 & 0.090 \\
TS      & 0.235 & 0.397       & 0.152 & 0.159          & 0.074 & 0.149 \\
ICT     & 0.193 & 0.310       & 0.254 & 0.603          & 0.410 & 0.277 \\
ICT+CPL & 0.061 & 0.023       & 0.042 & 0.070          & 0.138 & 0.071 \\
\hline
\end{tabular}
\caption{Symmetrized Kullback-Leibler divergence
of the arrival times$^*$ distributions computed on the original temporal network
and on the corresponding synthetic networks, for the conference datasets (HT09 and SFHH) and for the hospital dataset. 
KL indicates the divergence computed using the entire probability distribution,
while KL10+ corresponds to the divergence computed on the distribution tails only,
obtained by selecting arrival times$^*$ iarger than $10$.
}
\label{tabKL}
\end{table}

\subsection*{\small Determinants of the arrival time distribution}
 Using KL and KL10+ as guiding metrics we use the top-down approach to discard features
that are unimportant in reproducing the arrival time$^*$ distribution and to narrow down a set of necessary features.
We then use the bottom-up approach to find the features that are sufficient
to model the arrival time$^*$ distribution.

The Interval Shuffling (IS) and Link Shuffling (LS) procedures lead to
distributions of arrival times$^*$ similar to those of the empirical data.
This indicates that the causal structure of the temporal network has a small impact
on this distribution. Moreover,
LS does not preserve the specific assigment of cumulated contact durations $\omega_{AB}$ 
and number of contacts $n_{AB}$ to individual links (see Table~\ref{Table_Mod1}): 
we can therefore discard these 
as explanatory factors of the specific shape of the
arrival times$^*$ distribution.

According to Table~\ref{tabKL} and Fig.~\ref{fig:composite} (panels \textit{c-f})
the Time Shuffling (TS) procedure yields a very different distribution
for the conference datasets, and a different tail for the hospital data.
We know that, by design, the TS procedure does not preserve the distribution of inter-contact intervals,
which is directly related to the burstiness of contact activity.
The failure to adequately model the arrival times$^*$ distributions stems thus from this feature
and can be related to previous results~\cite{Vazquez:2007,karsai-2011-83}
showing that burstiness plays an important role in spreading phenomena. 
Indeed, the distribution of inter-contact interval durations for the synthetic networks
are quite different from those measured for the empirical networks
(not shown, established by comparing the KL divergences between the corresponding distributions). 
In the hospital case, this difference between empirical and synthetic inter-contact interval durations
is reduced, leading to the reduced difference in arrival times$^*$ distributions observed
in panel \textit{d} of Fig.~\ref{fig:composite} for the TS model.

However, panels \textit{c}-\textit{f} of Fig~\ref{fig:composite} also show
that the distributions of arrival times$^*$ obtained for the ICT model,
which is designed to preserve the distribution of inter-contact durations,
exhibit KL differences that are similar or even larger than those of the TS case discussed above.
The corresponding distributions in panels \textit{a} and {b} of Fig.~\ref{fig:composite}
are indeed much narrower than the ones obtained with the empirical dataset.
This shows that correctly reproducing the distribution of inter-contact durations
is not sufficient to adequately model the arrival time$^*$ distributions.
In order to achieve that, we need to add to the ICT model
the additional constraint of preserving the distribution of the number of contacts per link,
i.e., to use the ICT+CPL model (see Table~\ref{Table_Mod}). This model captures the essential features of the data
that are sufficient to reproduce the arrival time$^*$ patterns of the empirical data,
especially for the tail of the distributions, as shown in panels $e$ and $f$ of Fig.~\ref{fig:composite}.
We remark that this model is quite parsimonious, as it does not retain the topology of the empirical network
nor the distribution of contact durations or cumulated contact durations.

\subsection*{\small Activity-correlated classes of links}
\begin{figure}[!htbp]
\centering
\includegraphics[width=1.\textwidth, keepaspectratio]{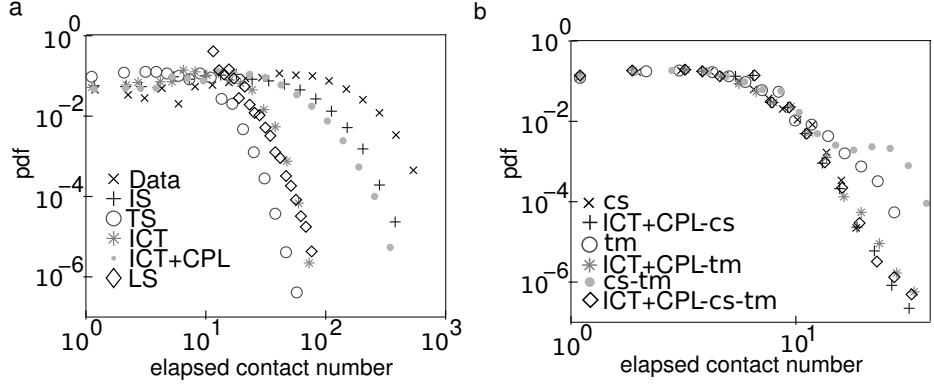}
\caption{a) Log-binned arrival time$^*$ distributions for the school dataset 
(original data, null models and generative models).
b) Log-binned arrival time$^*$ distributions for synthetic datasets 
generated by the toy model described in the main text,
together with the distributions obtained through the ICT+CPL model from  these synthetic datasets.
The synthetic datasets exhibit community structure (cs), temporal modulation (tm), or 
both  
(cs-tm).
Details are given in the Methods section.
$p_1=0.8,\ p_2=0.2$ for the cases with community structure,
and $p_1=0.5,\ p_2=0.5$ otherwise, see Methods.
\label{fig:dist}}
\end{figure}

 Despite the success of the ICT+CPL model for the conference and hospital datasets,
which we remark are quite different from one another, in the case of primary school data
none of these models yields a distribution of arrival times$^*$ close to that generated from the empirical data,
as shown in the panel \textit{a} of Fig.~\ref{fig:dist} and Table~\ref{KL-tabl-school}. 
In particular the IS and ICT+CPL models both yield similar, narrower distributions.

\begin{table}[!htbp]
\centering
\begin{tabular}{|l|l|l|l|l|l|l|}
\hline
models  & KL    & KL10+           \\
\hline
IS      & 2.113 & 2.483           \\
LS      & 3.040 & 3.504          \\
TS      & 4.455 & 5.361           \\
ICT     & 2.980 & 3.178           \\
ICT+CPL & 1.613 & 1.763       \\
\hline
\end{tabular}
\caption{Kullback-Leibler divergences between the arrival time$^*$ distributions
from the empirical school data and from each of the synthetic networks
based on the data. KL10+ indicates the divergence restricted to the tail of the distributions.
}
\label{KL-tabl-school}
\end{table}

Compared to the datasets considered in the previous sections,
the school dataset presents a few distinctive features.
In the conference cases individuals mix in a rather homogeneous way,
but most interactions occur at specific moments
typically corresponding to social activities such as coffee breaks~\cite{Cattuto2010,Isella:2011}.
In the hospital case, the interactions display characteristic role-dependent patterns,
but contacts are distributed rather homogeneously during the day~\cite{Isella:2011b}.
The primary school dataset, on the other hand, exhibits both a strong community structure
dictated by class membership, and correlated contact patterns across classes
determined by the schedule of social activities \cite{Stehle:2011}.
Contacts between children of different classes are possible during specific time intervals only,
and strongly correlated during such periods, because the school schedule controls class-based
activities rather than individual activities.

To tease apart the respective roles of community structure and correlated activity of link groups 
we  study the arrival time$^*$ distributions in the case of synthetic datasets 
exhibiting one or both of these features. These synthetic datasets are created 
using a toy model that generates temporal networks with tunable community structure (cs) and temporal
correlations in the activity of inter-community links. To this aim, we impose a temporal modulation (tm) 
in the activity of inter-community links: contact events on these links
can only occur during specific time intervals (see details in 
the Methods section). 
We subsequently compute arrival time$^*$ distributions 
for these synthetic datasets as well as for the corresponding ICT+CPL models.
Panel~\textit{b} of Figure~\ref{fig:dist} shows that, 
when activity-correlated classes of links are introduced, the arrival time$^*$ distributions
for the ICT+CPL case deviates significantly from that based on the corresponding
synthetic dataset.
This is similar to what we reported for the school data (even 
though the shape of the  arrival time$^*$ distribution is different) and occurs 
regardless of the presence (or lack thereof) or a community structure in the
synthetic dataset.
When the synthetic dataset displays a community structure but no correlations between the activity
of inter-community links,
the same arrival time$^*$ distributions are indeed observed for both the synthetic network 
and the corresponding ICT+CPL model.

\section*{Discussion}
The distribution of arrival times at various nodes of an epidemic process
unfolding over a temporal network, when measured in terms of activity clocks,
displays a behavior that is robust across very different settings and for different starting times of the process,
despite the intrinsic heterogeneities and non-stationarities of the temporal network.
The arrival time distribution expressed in terms of activity clocks thus represents an interesting tool
for investigating the structure of temporal networks beyond their surface features. 

The burstiness observed in many real-world networks,
indicated by a broad distribution of inter-event times,
is known to be an important feature of temporal networks
that influences dynamical processes taking place on them.
Here we have carried out an analysis based on empirical networks of human interactions,
measured in different social environments, and we have used 
suitably-designed null models and generative models for temporal networks
to show that the burstiness of inter-event sequences is not the only essential property
that needs to be retained when aiming at a realistic model of time-varying contact networks:
the heterogeneity of the number of contacts per individual link also plays a fundamental role
in determining the arrival times of the spreading process.
Our results show that, in fact, it is possible to design parsimonious generative models
of temporal networks, such as the ICT+CPL model, based on just the distribution of inter-event interval durations
and on the distribution of number of contacts per link. The ensuing synthetic temporal networks
adequately model the arrival time distributions of real-world networks measured in diverse settings.

Interestingly, the behavior of the arrival time distribution expressed in terms of activity clocks
is sensitive to complex features of the temporal network data such as the presence
of activity-correlated classes of links, as exemplified by the case of the school temporal network,
where the interplay of the community structure induced by classes and of correlated activity
pattens due to schedule activities creates rich temporal structures in the data.
We have shown that the presence of classes of links that are only active
in a correlated fashion during specific time windows
has an impact on the spreading time distribution and breaks down the ability to use
parsimonious models such as the ICT+CPL one.
Activity-correlated classes of links, which are arguably common in many real-world social systems,
are difficult to uncover on the basis of simple statistical observables for the temporal network,
and their impact on the dynamics of spreading process calls for more research.
We have shown that arrival time distributions based on activity clocks are a precious tool
in this respect as they have the ability to indicate the presence of such complex structures and correlations.
Simple generative models, such as the ICT+CPL model, cannot possibly
account for these complex structures and should thus be enriched, when necessary,
by introducing additional features such as classes of links with correlated and temporally-localized activity.
Here we have shown that toy models that minimally incorporate such features
yield deviations in the arrival time patterns similar to those observed for the school temporal network.

Overall, our results call for more work in the direction
of both detecting and modeling complex temporal-topological structures in time-varying networks.
Similarly, more work is needed to design minimal generative models that incorporate
realistic features found in empirical data from real-world scenarios.

\section*{Methods}

\subsection*{\small Definition of null and generative models}
Here we describe the different shuffling procedures and generative models introduced in the main text.
For the top-down approach, we start from the empirical temporal networks,
on which we apply the following shuffling procedures:
\begin{description}
\item [Interval Shuffling (IS):]
The sequence of contact and inter-contact intervals
of each link is randomly shuffled. The original contact durations and inter-contact durations
are thus preserved. Given a link $(a,b)$ with $n$ contact events,
let us denote the contact intervals by ${(s_0,e_0), (s_1,e_1), ..., (s_n,e_n)}$.
The set of contact durations is thus given by ${(e_0-s_0), (e_1-s_1), ..., (e_n-s_n)}$,
and the set of inter-contact durations is ${(s_1-e_0), (s_2-e_1), ..., (s_n-e_{n-1})}$.
We create a synthetic timeline for the link $(a,b)$ by randomly shuffling
the sequence of contact and inter-contact intervals, and then we randomly and uniformly translate
the starting time $s'_0$ of the link's new timeline within the remaining time interval $T-(e_n-s_0)$, where $T$ is the full dataset time interval.
Consistently, links with $n=1$ contact events have an empty set of inter-contact times
and the single contact interval is simply randomly displaced in time.
\item [Link Shuffling (LS):]
Whole single-link event sequences are randomly
exchanged between randomly chosen link pairs.
Event-event and weight-topology correlations are destroyed.
\item [Time Shuffling (TS):]
Time intervals of the whole original contact sequence are randomly shuffled
and reallocated randomly to each link retaining the distribution of the number of contacts per link of the original dataset.
Temporal correlations are destroyed. The resulting shuffled network is built with a condition
of no intersection between contact intervals in the same link.
\end{description}

In the case of the generative models, we start by creating a static random network
with approximately the same degree distribution, the same number of nodes,
and the same number of links as the empirical network we want to study.
Then, we build a temporal network by associating with each link a sequence of contact events,
according to the following strategies:
\begin{description}
\item [ICT:] For each link we set the number of contacts
per link equal to the average number of contacts per link of the original data.
Each of these contacts is then generated with a duration equal to the average contact duration
observed in the empirical data. The time between contact events is set by sampling with replacement
the distribution of inter-event times measured in the data.
\item [ICT+CPL:] The ICT+CPL model is based on the ICT model described above,
with the additional constraint that for each link the number of contacts is not constant,
but is set by sampling with replacement the distribution of the number of contacts per link of the empirical data.
\end{description}

\subsection*{\small Symmetrized Kullback-Leibler divergence}
The symmetrized Kullback-Leibler divergence is defined as:
\begin{equation}
    DIV^{s}_{\mathrm{KL}}(M\|D) = \frac{1}{2} \left( \sum_i M(i) \log \frac{M(i)}{D(i)} + \sum_i D(i) \log \frac{D(i)}{M(i)} \right) \, ,
 \end{equation}
where $D(i)$ is to the distribution of (integer-valued) arrival times$^*$ in the empirical data
and $M(i)$ is the distribution yielded by the models.
To assess the stochastic variability range of the Kullback-Leibler divergence,
we generate several realizations of each null model or generative model,
we compute the divergence between the distribution yielded by each realization and that of the original data,
and we show a box plot summarizing the resulting values.

\subsection*{\small Definition of a toy model with activity-correlated link classes}
In order to understand which features of the school data
make the distribution of arrival times$^*$ not reproducible by the synthetic networks
of the ICT+CPL model, we introduce a toy generative model
that produces temporal networks with some key features of the original school network,
namely the community structure and the synchronization of the activity/inactivity
patterns of some groups of links.
We start by building a static network with a simple two-community structure:
we consider $N$ nodes and divide them into two groups of equal size.
Within each group, two nodes are  linked with a probability $p_1$.
Nodes across the two communities are linked with a probability $p_2 \leq p_1$
(the case $p_1 = p_2$ yields a random graph without community structure).
This procedure defines the topological structure of the network.
We build the temporal network by associating with each link a sequence of contact events.
These activity sequences are all generated by sampling a Poisson process with a rate $\lambda=0.0056 \, s^{-1}$,
which was chosen to yield an average number of contacts per link of the same order of the school data
over the same global time $T \simeq 100,000 \, s$.
For the cross-community links we then remove all events outside
of the interval $[T/2 (1 - \delta), T/2 (1 +  \delta)]$.
This last condition introduces a temporal modulation for the inter-community links,
which are only active in the above time window. In the limit $\delta \rightarrow 1$ we recover the non-modulated case.

\section*{Acknowledgements}
CC and AB are partly supported by the EU FET project MULTIPLEX (grant number 317532).

%

\section*{Author Contributions}
L.G and A.P contributed equally to the work. L.G., A.P, C.C and A.B
designed the study.
L.G and A.P carried out the data analysis and performed the simulations
L.G., A.P, C.C and A.B wrote and reviewed the manuscript.
\section*{Additional Information}
The author(s) declare no competing financial interests.

%

\pagebreak
\section*{Supplementary Information}
\subsection*{\small Description of the data}
The data have been collected using the 
infrastructure developed in the framework of 
the SocioPatterns collaboration (http://www.sociopatterns.org).
This infrastructure is based on
Radio Frequency Identification Devices (RFID) that record the face-to-face contacts (distance of about $1$ meter) with a
time resolution of $20$ seconds. The data come from $4$ deployments, in a 
hospital, in a conference (HT09), in a congress (SFHH) and in a primary school. Some of the features of
the data sets are listed in Table \ref{TableData}.

\begin{table}[!htbp]
 \centering
\begin{tabular}{|l|c|c|c|}
\hline
deployment  & number of participants & duration (days) \\
\hline
HT09 conference & $112$ & $3$ \\
SFHH congress & $415$ & $2$ \\
hospital      & $174$ & $8$ \\
school    & $251$ & $2$ \\
\hline
\end{tabular}
\caption{Characteristics of the datasets.}
\label{TableData}
\end{table}

\subsection*{\small Comparison of generated and empirical and synthetic distributions}

\begin{figure}[!htbp]
\centering
\includegraphics[width=\textwidth,keepaspectratio]{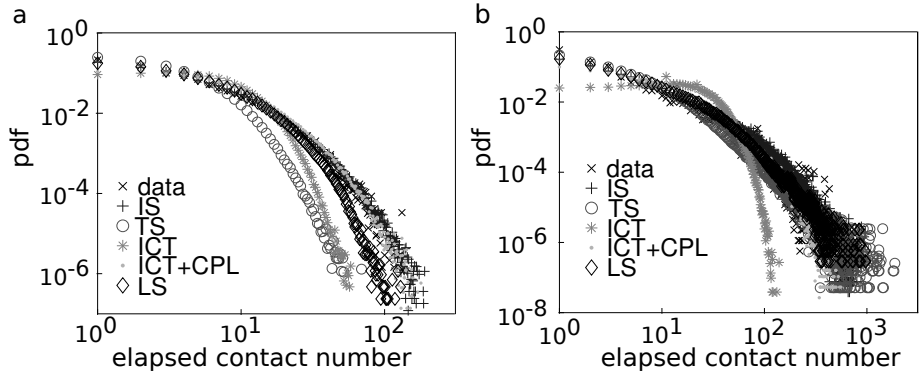}
\caption{
Probability distributions of arrival times$^*$
for the conference dataset (HT09, panel a) and the hospital data (HOSP, panel b).
In each panel, IS, LS, TS, ICT, ICT+CPL stand respectively
for Interval Shuffling, Link Shuffling, Time Shuffling,
Inter-Contact time model, and Inter-Contact time plus Contacts-per-Link model.
``Data'' indicates the distribution of arrival times$^*$
obtained by simulating an SI process over the empirical temporal network
($200$ realizations with random starting times for each node of the network taken as seed of the epidemics).
For each model, we consider $20$ different realizations of the temporal network.
For each of these realizations we run $20$ different SI epidemics, each with a different random starting time.
The arrival times$^*$ (top row) for all those runs are aggregated to yield the reported
distributions.
\label{fig:composite_notbin}
}
\end{figure}

\begin{figure}[!htbp]
\begin{center}
\includegraphics[width=6cm] {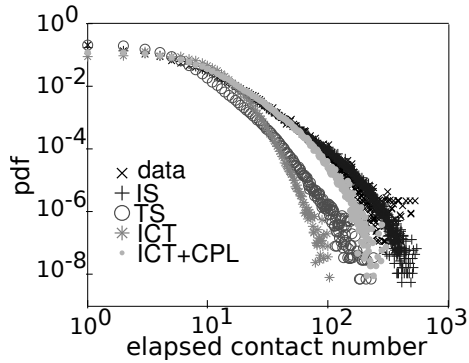}
\caption{Distribution of arrival times$^*$ for the SFHH data. IS, TS, ICT, ICT+CPL 
respectively stand for Interval Shuffling, Time Shuffling, Inter-Contact Time model, Inter-Contact Time plus
Contact-Per-Link model.}
 \label{sfhh}
\end{center}
\end{figure}

\begin{figure}[!htbp]
\centering
\subfigure[SFHH]{
	\label{fig:kldiv-sfhh-1}
	\includegraphics[width=0.45\textwidth, keepaspectratio]{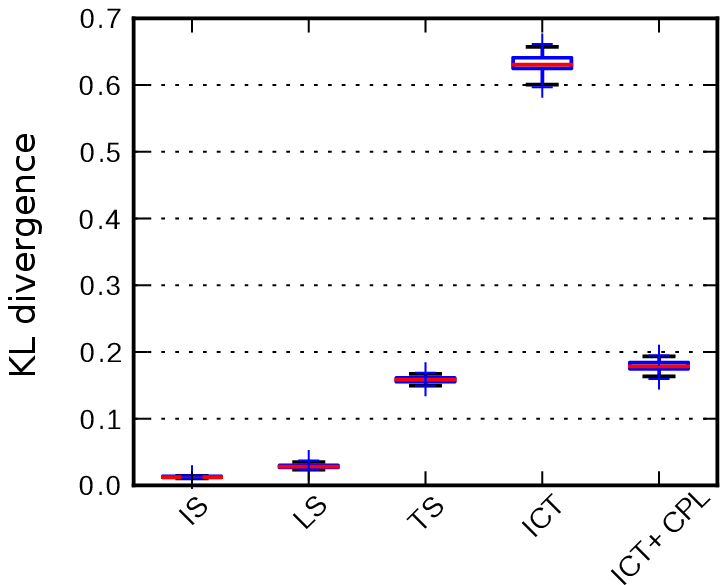}
}
\subfigure[SFHH 10+]{
	\label{fig:kldiv-sfhh-2}
	\includegraphics[width=0.45\textwidth, keepaspectratio]{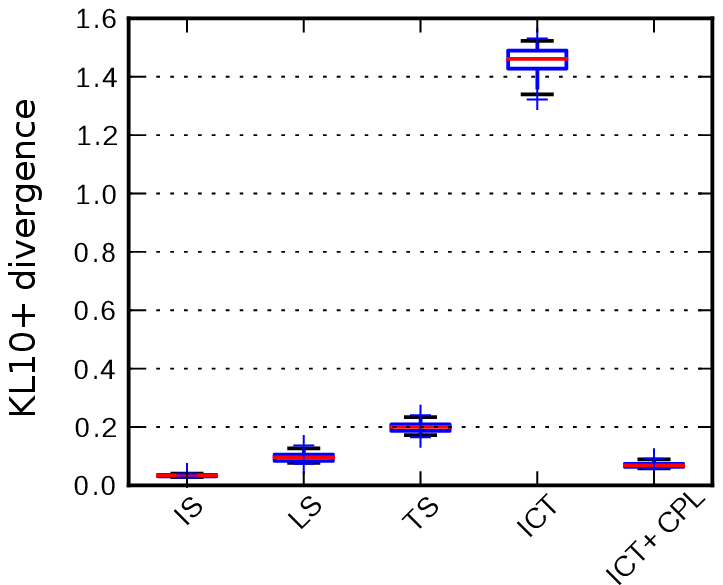}
}
\caption[]{Kullback-Leibler divergence for the SFHH data. The results
  come from 20 different realizations of each model, and, for each realization, 
  for 20 different starting times. The arrival times$^*$ for each
  realization and each starting time are measured and the resulting distribution is
  compared to the distribution generated by the original data. In the
  first plot, we keep all generated values, while in the second plot
  we filter out all values lower than $10$ in order to show the KL
  divergence restricted to the tail of the distributions. The box extends from the lower to upper
  quartile values of the data, with a line at the median.  The
  whiskers extend correspond to the $95\%$ confidence interval. The points
  past the end of the whiskers show the outliers.}
\label{fig:kldiv-sfhh}
\end{figure}

Here we present the results of the SI process simulated respectively
on the conference, on the hospital (Fig.\ref{fig:composite_notbin})  and on the SFHH data and on all the corresponding synthetic temporal networks created by the null models
and the generative models. Figures~\ref{sfhh} and \ref{fig:kldiv-sfhh} confirm 
the efficiency of the ICT+CPL model to reproduce the arrival time$^*$ distribution in the SFHH case.
\pagebreak

\begin{figure}[!htbp]
\centering
\includegraphics[width=1.\textwidth, keepaspectratio]{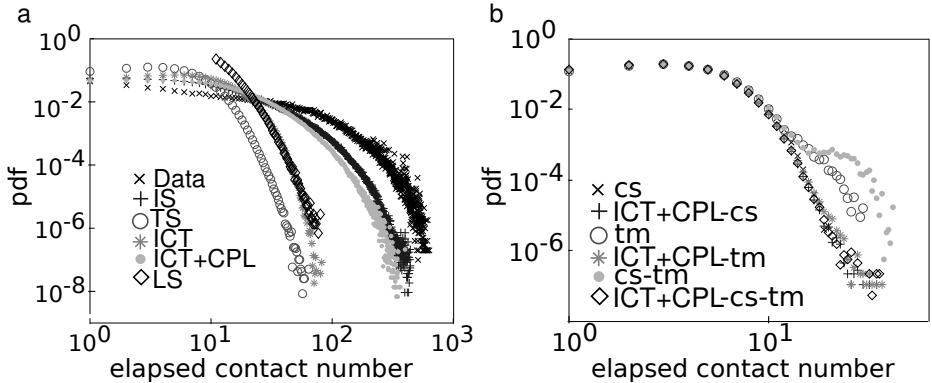}
\caption{a) arrival time$^*$ distributions for the school dataset (original data, null models and generative models).
b) arrival time$^*$ distributions for different temporal networks (cs,tm,cs-tm) generated by the toy model described in the main text,
together with the distributions obtained through the ICT+CPL model based on the toy model networks.
cs) with community structure only; tm) with temporal modulation only; cs-tm) with both community structure and temporal modulation.
Details are given in the Methods section.
$p_1=0.8,\ p_2=0.2$ for the cases with community structure,
and $p_1=0.5,\ p_2=0.5$ otherwise, see Methods.
All networks generated by the toy model have approximately the same number of links.
\label{fig:dist_notbin}}
\end{figure}

Panel \textit{a} of Fig.\ref{fig:dist_notbin} displays the results of the SI process simulated on the school data and on all the corresponding synthetic temporal networks created by the null models
and the generative models .
On panel \textit{b} of Fig.\ref{fig:dist_notbin}, we present together the results of the SI process simulated on the synthetic networks generated by the toy model described in the Methods section and on the corresponding temporal networks created by the ICT+CPL model.

\end{document}